\documentclass[twocolumn, showpacs,preprintnumbers,amsmath,amssymb]{revtex4}
\usepackage{amssymb}
\usepackage{xcolor}
\usepackage{graphicx}
\usepackage{dcolumn}
\usepackage{bm}
\usepackage{multirow}
\usepackage{booktabs}
\usepackage{natbib}
\usepackage{soul}
\usepackage{amssymb}
\usepackage{amsmath}	
\usepackage{color}
\usepackage{cases}

\newcommand{\etal}{\textit{et al.}\ }

\newcommand{\Rmat}{{\it R}-matrix }

\newcommand{\npone}{\mbox{$N+1$}}
\newcommand{\bohr}{$\,a_0$}

\begin{document}
\preprint{APS/123-QED}

\title{Rydberg states and new resonant states of the imidogen molecule NH: pathways for nitrogen release}

\author{Raju Ghosh$^{1,2}$}\email[]{rajughosh152@gmail.com}
\author{Binayak Samaddar Choudhuryt$^{2}$}\email[]{binayak12@yahoo.co.in}
\author{J\'anos Zsolt Mezei$^{3,4}$}\email[]{mezei.zsolt@atomki.hu}
\author{Ioan F. Schneider$^{4,5}$}\email[]{ioan.schneider@univ-lehavre.fr}
\author{Nicolina Pop$^{6}$}\email[]{nicolina.pop@upt.ro}
\author{Kalyan Chakrabarti$^{7}$}\email[]{kkch.atmol@gmail.com}
\affiliation{$^{1}$Dept. of Mathematics, Sukumar Sengupta Mahavidyalaya, Keshpur, Paschim Medinipur 721150, India }%
\affiliation{$^{2}$Dept. of Mathematics, Indian Institute of Engineering Science and Technology, Shibpur, Howrah 711103, India}%
\affiliation{$^{3}$HUN-REN Institute for Nuclear Research (ATOMKI), H-4001 Debrecen, Hungary}%
\affiliation{$^{4}$LOMC CNRS-UMR8112, University Le Havre Normandy, Le Havre 76000, France}%
\affiliation{$^{6}$LAC-UMR9188, CNRS Universit\'e Paris-Saclay, F-91405 Orsay, France}%
\affiliation{$^{7}$Dept. of Mathematics, Scottish Church College, Kolkata 700 006, India}%
\date{\today}

\begin{abstract}
Neutral resonant states of molecules play a very important role in the dissociation dynamics and other electronic processes that occur via intermediate capture into these states. With the goal of identifying resonant states, and their corresponding widths, of the imidogen molecule NH as a function of internuclear distance, we have performed detailed \Rmat  calculations on the e + NH$^+$ system. In a previous work, we had identified bound states of NH and Feshbach resonances in the e + NH$^+$ system at a single geometry, namely the NH$^+$ equilibrium $R_e=2.0205$\bohr. Here we present a much more detailed work by repeating the calculation on over 60 internuclear distances to obtain the corresponding potential energy curves. The bound states for nine symmetries have been detailed many of which, particularly the singlet states, were never studied before. Several resonant states of different symmetries, which were unknown until now, have been systematically identified and their widths calculated in the present work, which proved much more challenging due to presence of many avoided crossings. It is hoped that the bound and the new resonant states obtained here will open up other molecular dynamics studies, since for several dissociative processes, although experimental data existed for more than a decade, these are still uncorroborated due to absence of molecular data, and hence subsequent theoretical calculations.
\end{abstract}


\maketitle

\section{Introduction}

Controlling heat load on the plasma facing components (PFCs) is a major issue in large magnetically controlled fusion devices like ITER, and is needed to protect the PFCs from thermal damage. Injection of impurities like nitrogen, and noble gases like neon and argon are being increasingly used and tested for this purpose \cite{Casali22,Eldon22}. Injected impurity gases can interact with the plasma giving rise to many other species of molecules and ions. In particular, since the plasma contains significant quantities of H, D and T from the fuel, significant amounts of NH, NH$^+$ and their isotopologues are expected for the case of injected nitrogen. Detailed collision data in the form of cross sections and rates for these molecules and ions would therefore be required as components for plasma flow modeling. On the other hand, many electron driven processes such as dissociative recombination, dissociative excitation, dissociative electron attachment and resonant ro-vibrational excitation undergo via resonant states and hence molecular data sets related to these resonant states, which are substantially difficult to obtain, are required to initiate theoretical studies on these processes.

On the other hand, these species are also relevant for the interstellar medium (ISM).  Indeed, NH was first observed in the interstellar medium at the beginning of the 1990s by Meyer and Roth \cite{Meyer91} and since then it was found in different protostars \cite{Benz10} and photon dominated regions\cite{Fuente10}. It is the minority final product of the electron impact dissociative recombination of N$_2$H$^+$, another key member of the nitrogen astrochemistry of ISM \cite{Mezei23}.

A theoretical investigation of the ortho-para chemistry of ammonia in the cold ISM was done by Rist et al.\cite{Rist13} They calculated separate branching ratios for the hydrogen abstractions and DR reactions leading to the ortho and para forms of nitrogen hydrides, taking into account of nuclear spin selection rules.  The chemistry of nitrogen in dense regions of the ISM that are efficiently protected from ultraviolet photons by the dust and molecular hydrogen was treated by Le Gal et al.\cite{LeGal14} Moreover, the imidogen radical NH and its ion NH$^+$ have also been reported in comets as early as 1941 \cite{Swings41, Lecointre10,Amero04}. The abundance of ammonia in tails of some comets may be explained through the photodestruction of NH which subsequently leads to the formation of NH$_3$.

The presence of ammonia in dense clouds is suggested by  Le Bourlot~\cite{Bourlot91} by means  the gas-phase synthesis through the N$^+$ + H$_2$ reaction, followed by hydrogen abstractions and DR reactions, enough to reproduce the observed amounts. Consequently, a detailed picture of the molecular states of NH are needed for studies on photodissociation and other collision processes that initiate the nitrogen chemistry.

Our previous work on NH$^+$ \cite{Ghosh22} was focused on obtaining electron collision cross sections. In that work, we had also performed some of the groundwork for obtaining the molecular states of NH and calculated some of them at a single geometry, namely the NH$^+$ equilibrium $R_e=2.0205$\bohr. Our calculations showed many Feshbach resonances which may possibly build up resonant (dissociative) states that could be routes for release of N atoms and N$^+$ ions.

Although the major motivation for the present work comes from the presence of NH and NH$^+$ in fusion devices from injected nitrogen impurities, and the requirement of molecular data to initiate collision studies for much needed collision cross sections and rates, the above discussions show that these molecular states are also highly significant in the context of astrophysics.

In the present work, we report the identification and the characterisation  of highly-excited bound states of NH close to its ionization threshold, revealed by the R-matrix modeling of the scattering of an electron on the NH$^+$ cation. These Rydberg states are characterized by their quantum defects, or equivalently, their effective quantum numbers, which we obtain  systematically. We also detect the major dissociative states of NH of $^1\Sigma^+$, $^1\Sigma^-$, $^1\Pi$ and $^3\Sigma^+$ symmetries that are identified  from e+NH$^+$ resonances, and we obtain the corresponding autoionization widths which characterize the Rydberg-valence interactions.

\section{Theoretical Method}{\label{sec:theory}}

Developed over several decades, the molecular \Rmat method is highly advanced and now represents the state of the art in electron-molecule collision calculations. Although primarily a scattering formalism, as outlined later, it can also be used to calculate electronic states from a collision calculation. The method is described in our earlier works (see for example \cite{Mukherjee24}) and, in much more detail, in the review by Tennyson \cite{Tennyson10}. To avoid repetition we give only its outline below.

Assuming collision of an electron with an $N$ electron target ($N=7$ here for NH$^+$), the \Rmat method first separates the configuration space into an inner region, a sphere of radius $a$ (the \Rmat radius) centered on the target center of mass, and an outer region exterior to it. In the inner region, the wave function of the scattered electron+target system with (\npone)-electrons and symmetry $\Upsilon$ is written as a close coupling (CC) expansion,
\begin{eqnarray}\label{eq:cc}
\Psi_k^\Upsilon = &&\mathcal{A} \sum_{i,j,k} a_{ijk}\; \Phi_i(x_1,\cdots, x_N)\; u_{ij}(x_{N+1}) + \nonumber\\&&\sum_i b_{ik}\; \chi_i(x_1,\cdots, x_{N+1}),
\end{eqnarray}
where in Eq. (\ref{eq:cc}), $\mathcal{A}$ is an antisymmetrisation operator,  $\Phi_i(x_1,\cdots, x_N)$ is the $i^{th}$ target state wave function and $u_{ij}(x_{N+1})$ are continuum functions representing the scattered electron and the $k^{th}$ component of $\Psi^\Upsilon$ is denoted by $\Psi_k^\Upsilon$. The continuum functions $u_{ij}$ chosen, depend on the target charge and can be either Coulomb functions for positively charged targets, as in the present case, or spherical Bessel functions for neutral targets.
The functions $\chi_i(x_1,\cdots, x_{N+1})$ are called $L^2$ functions since they are square integrable and are used to account for the polarisation of the target in presence of the incident electron. There is no unique prescription for the $L^2$ functions and these are constructed by allowing the incident electron to occupy the target molecular orbitals (MOs). The coefficients $a_{ijk}$ and $b_i$ are determined variationally to get $\Psi^\Upsilon$. This inner region solution is the most computationally intensive part, however, it needs to be done only once for all scattering energies.

From the inner region wave function $\Psi^\Upsilon$ an \Rmat is then built at the boundary of the \Rmat sphere at a given scattering energy $E$. The \Rmat $\mathbf{R}(r,E)$ is connected to the logarithmic derivative of the radial part of $\Psi^\Upsilon$ by
\begin{equation}
 \mathbf{R}(r,E) = \mathbf{f}(r)[r\; \mathbf{f}'(r)]^{-1},
\end{equation} 
where $\mathbf{f}(r)$ is the radial wave function and $r$ the radial coordinate \cite{Tennyson10}. For scattering or bound state solutions, the radial wave function $\mathbf{f}(r)$, or equivalently the \Rmat, must satisfy certain asymptotic boundary conditions (see Tennyson (2010)\cite{Tennyson10} for example). To make $\mathbf{f}(r)$ satisfy the correct asymptotic boundary conditions, the \Rmat is propagated numerically to a suitable asymptotic distance $R_{asy}$ and finally matched with known asymptotic solutions expressed in terms of a Gailitis expansion \cite{Noble84}. This matching gives the $\mathbf{K}$-matrix which contains all scattering information. As discussed below, the resonances and their widths can be found from the $\mathbf{K}$-matrix, while the bound states are obtained by matching the radial wave function to exponentially decreasing functions\cite{Tennyson10,Sarpal91}.

\section{Calculations}
\subsection{Target calculations}\label{sec:targ} 

A highly detailed exposition of our target calculations is available in our earlier work on \cite{Ghosh22}, where the calculations were done at a single geometry, namely the NH$^+$ equilibrium $R_e=2.0205$\bohr. There, we had performed extensive tests to find suitable target and scattering models in the framework of the \Rmat method to obtain various collision cross sections for electron impact on the NH$^+$ target. Moreover, we had also, at the NH$^+$ equilibrium, identified many Feshbach resonances in the e+NH$^+$ system and bound states of the NH radical. For the benefit of the readers, we briefly outline below the details of the target and scattering calculations.

\subsubsection{NH$^+$ target model}\label{sec:target}
We have used the Slater type orbitals (STOs) of Cade and Huo \cite{Cade67,Cade73} to make an SCF calculation on the  X~$^2\Pi$ ground state of NH$^+$ at the equilibrium $R_e=2.0205$\bohr. Since it is difficult to get accurate energies of open shell molecules, we then obtained a set of $^2\Pi$ natural orbitals (NOs) from a complete active space singles and doubles (CAS+SD) calculation on the NH$^+$ X~$^2\Pi$ state using the SCF orbitals. Subsequently we used these NOs in a configuration interaction (CI) calculation. For the target states $\Phi_i$ in Eq. (\ref{eq:cc}), a CAS-CI wave function was used and after many tests the target model $(1\sigma)^2(2\sigma-8\sigma, 1\pi-3\pi, 1\delta)^5$ was selected as it gave vertical excitation energies (VEEs) in excellent agreement with the corresponding VEEs of the multi-reference single and double configuration interaction (MRSDCI) calculations of Kusunoki \etal\cite{Kusunoki86} and Amero \etal\cite{Amero05} (see Table 1 of our previous work \cite{Ghosh22}). This target model was used in all subsequent scattering calculations.

\begin{figure}[t]
\centering
\includegraphics[width=0.9\columnwidth]{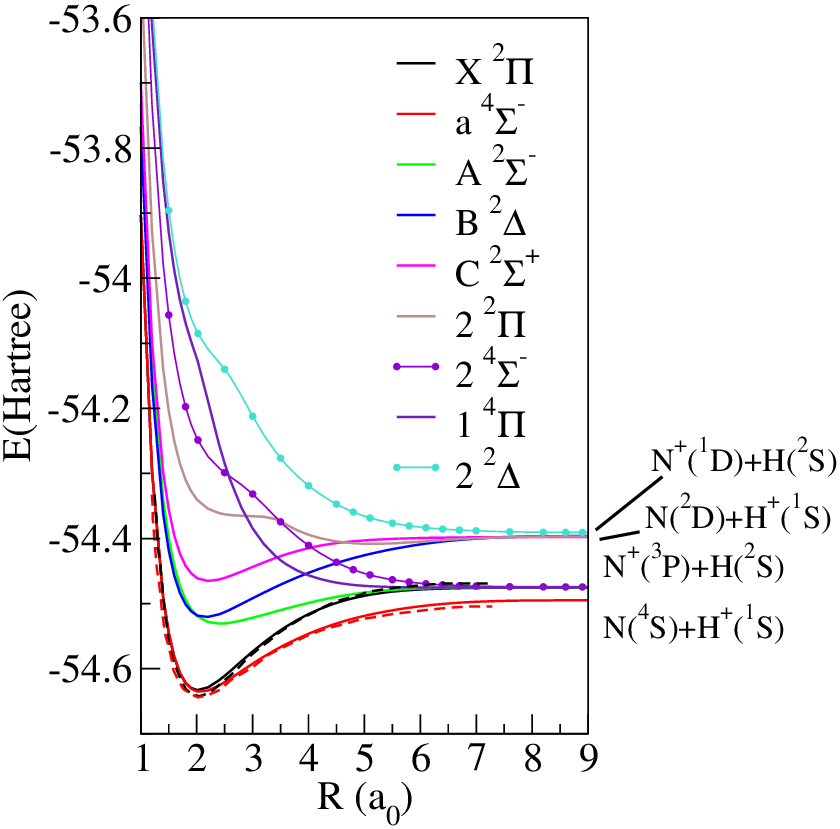}
\caption{Potential energy curves for the first 9 low lying states of the target NH$^+$ ion. The ground state (black curve) is X~$^2\Pi$. The dashed curves are the X~$^2\Pi$ (dashed black) and the a~$^4\Sigma^-$ (dashed red) states obtained by Amero and V\'azquez \cite{Amero04}.\label{fig:targ}}
\end{figure}

Figure \ref{fig:targ} shows the first 9 NH$^+$ target states. As obtained in earlier calculations (see \cite{Amero04} for example), near the ground state equilibrium, the X~$^2\Pi$ ground state and the a~$^4\Sigma^-$ state are very close. The 2~$^4\Sigma^-$, 1~$^4\Pi$ and the 2~$^2\Delta$ states are repulsive. For a comparison, we have also shown the X~$^2\Pi$ ground state and the a~$^4\Sigma^-$ state of Amero and V\'azquez \cite{Amero04} as dashed lines, which compare excellently with the corresponding PECs of our calculations.

\subsection{Scattering calculations: bound and resonant states}
Scattering calculations were performed with the same 12 NOs (namely $8 \sigma$, $3 \pi$ and $1 \delta$) obtained as outlined above and the target model described in as in Section \ref{sec:target}. Referring to Eq. (\ref{eq:cc}), the scattered electron in the continuum was represented by continuum functions $u_{ij}$ which were obtained as a partial wave expansion about the target center of mass, retaining those partial waves with $l\le 6$ and $m\le 2$. The $l$ and $m$ values were carried over from our previous work \cite{Ghosh22}, where they were chosen to obtain convergence in cross sections, bound state energies and resonance positions. Since the target was positively charged, the functions $u_{ij}$ were chosen to be Coulomb functions, which were numerically generated by solving the radial Coulomb equation retaining solutions below 10 Ryd. An \Rmat radius of 11\bohr~was chosen for internuclear distances smaller than the NH$^+$ equilibrium $R_e$, however, it needed to be progressively increased for calculations with larger internuclear distances to ensure confinement of the stretched target within the \Rmat box going up to 16.5\bohr~for $R=9.0$\bohr.

Extensive tests were made on different scattering models by varying the number and nature of the of target states for each symmetry of the inner region wave function $\Psi_k^\Upsilon$ in the close coupling expansion Eq. (\ref{eq:cc}). The final scattering models were chosen so that energy of the X~$^3\Sigma^-$ ground state of the NH was lowest, and the VEEs of next 30 excited states of NH were in good agreement with existing results \cite{Owono07,Rajvanshi10,Goldfield87}, the details of all these can be found in our earlier work \cite{Ghosh22} and is omitted here for brevity. 

The scattering calculations were repeated over a moderately dense grid of 61 geometries for $R=1$\bohr~ - 9\bohr~for the symmetries $^1\Sigma^+, ^1\Sigma^-, ^1\Pi, ^1\Delta, ^3\Sigma^+, ^3\Sigma^-, ^3\Pi, ^3\Delta, ^5\Sigma^-$ and $^5\Pi$, where the PECs of the neutral states have reached their asymptotic behavior.

\subsubsection{Bound states of NH}
In the \Rmat formalism, the bound states can be found from a scattering calculation, provided that the (\npone)-electron wave function of the e+NH$^+$ system  satisfies the correct asymptotic boundary conditions. For bound states, it is known that the wave functions should exponentially decrease to zero asymptotically. With this prescription, the \Rmat built on the boundary was propagated to 50\bohr~in a potential induced by the diagonal and off diagonal dipole and quadrupole moments of NH$^+$. The bound states were found as negative energy solutions of the outer region problem, and were located by searching over a non linear quantum defect grid (see for example Sarpal et al\cite{Sarpal91}).

\subsubsection{Resonances}
Resonances, which are transient quantum states, are obtained by temporary capture of the incident electron by the target. Resonances occur numerously in electron collision with positive ions and are characterised by their energy $E^r$ (position), and the width $\Gamma^r$ which is inversely proportional to their lifetime $\tau$. While more discussion on these states appear below, we give here the outline of the procedure to obtain resonance parameters. 
For our calculations the \Rmat was propagated to 70\bohr~and then matched with asymptotic functions obtained from a Gailitis expansion\cite{Noble84} to obtain the $\mathbf{K}$-matrix. 
The eigenphase sum $\delta(E)$ was then obtained from the $\mathbf{K}$-matrix and is defined from the diagonal elements $K_{ii}$,
\begin{equation}
    \delta(E)=\sum_i \tan^{-1}(K_{ii}).
\end{equation}
It is known that near a resonance, the second derivative of $\delta(E)$ undergoes a characteristic jump by $\pi$\cite{Tennyson84}. The resonance energies $E_i^r$ were obtained by numerically locating the energy positions at which the changes in sign occur, and finally their widths $\Gamma_i^r$ were obtained by fitting to a Breit-Wigner profile\cite{Tennyson84},
\begin{equation}
    \delta(E)=\sum_i \tan^{-1}\Big[ \frac{\Gamma_i^{r}}{E-E_i^r} \Big] + b(E)
\end{equation}
where $b(E)$ is the background, which is usually chosen to be a linear or a quadratic polynomial to represent the underlying trend of $\delta(E)$ near the resonance.

\begin{table}[t]
\small
\caption{Quantum defects (QD) of some of the singlet and triplet states of NH at R$_e$ = 2.0205\bohr. Shown also, in the last column, are the vertical excitation energies from the NH X~$^3\Sigma^-$ ground state. }
\label{tab:QD}
\begin{tabular} {llccc} 
\hline
State & Rydberg& QD & QD & VEE\\ 
& orbital & (Present) & (Expt.) & (eV)$^d$ \\
\hline
$1(b) ^1\Sigma^+$&  2p$\pi$ &  0.856  & 0.881$^b$ & 2.86\\
$2(h) ^1\Sigma^+$& 3p$\pi$ & 0.566 & 0.614$^b$ & 10.97\\
$3 ^1\Sigma^+$& 3d$\pi$ & $-0.006$  & & 11.76\\
$4 ^1\Sigma^+$& 4p$\pi$ & 0.561  & & 12.12\\
$5 ^1\Sigma^+$& 5p$\pi$ & 0.561  & & 12.58\\\\

$1 ^1\Pi$  &2p$\sigma$ &  0.661   & & 5.69\\
$3 (f) ^1\Pi$  &3s$\sigma^e$ &  0.718  & 0.788$^b$ & 10.65\\
$4^1\Pi$   &3p$\sigma$ &  0.320 & & 11.37\\
$5^1\Pi$   &3d$\sigma$ &  0.069 & & 11.68\\\\

$2 (g) ^1\Delta$  & 3p$\pi$   & 0.634 & 0.696$^b$ & 10.84 \\
$3^1\Delta$  & 3d$\pi$   & 0.018  & & 11.74\\
$4^1\Delta$  & 4p$\pi$   & 0.674 & & 12.04\\
$5^1\Delta$  & 4d$\pi$   & 0.022 & & 12.41\\\\

$1^1\Sigma^-$ &3p$\pi$  &  0.807 & & 10.44\\
$2^1\Sigma^-$ &3d$\pi$  &  0.112 & & 11.63\\
$3^1\Sigma^-$ &4p$\pi$  & 0.778  & & 11.96\\
$4^1\Sigma^-$ & 4d$\pi$ & 0.069$^a$  & & 12.49\\

\hline
$1^3\Sigma^+$ & 3p$\pi$ & 0.729  & & 10.63\\
$2^3\Sigma^+$ & 3d$\pi$ & 0.013 & & 11.74\\
$3^3\Sigma^+$ &4p$\pi$  & 0.720  & & 12.00\\
$4^3\Sigma^+$ &4d$\pi$  & 0.026   & & 12.40\\
$5^3\Sigma^+$ &4f$\pi$  & $-0.004$  & & 12.42\\\\

$1 (A) ^3\Pi$   & 2s$\sigma$ & 0.791 & & 3.88\\
$3(B) ^3\Pi$ & 3s$\sigma$ & 0.727  & 0.820$^c$ & 10.57\\
$4(D)^3\Pi$ & 3p$\sigma$ & 0.440 & 0.640$^c$ & 11.12\\
$5^3\Pi$ & 3d$\sigma$ & $-0.059$ & & 11.75\\\\

$3^3\Sigma^-$ & 3p$\pi$ & 0.768  & & 10.47\\
$5^3\Sigma^-$ &3d$\pi$ &0.192$^a$  & & 11.52\\
\hline
\end{tabular}\\
$^a$ At $R=2.2$\bohr.\\
$^b$ Calculated from values given by Johnson III and Hudgens.\cite{Jhonson90}\\
$^c$ Quoted from Clement \etal\cite{Clement92b}\\
$^d$ Quoted from our previous work, Ghosh et al.\cite{Ghosh22}\\
$^e$ Labeled $3p\sigma$ by Johnson III and Hudgens\cite{Jhonson90}.
\end{table}

\begin{figure}[t]
\centering
\includegraphics[width=0.85\columnwidth]{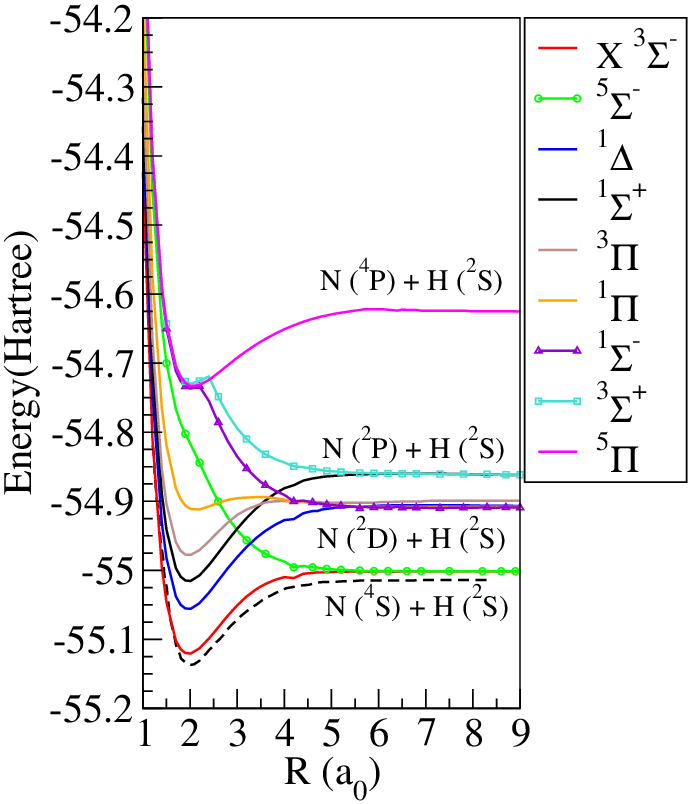}
\caption{Potential energy curves for the first 9 low lying states of the NH radical. The ground state (bottom most continuous curves) is X~$^3\Sigma^-$. The lowest dashed curve is the X~$^3\Sigma^-$ ground state of Owono et al \cite{Owono07}.}
\label{fig:bound}
\end{figure}

\section{Results and discussion}

\begin{figure*}[t]
\centering
\includegraphics[width=0.4\textwidth]{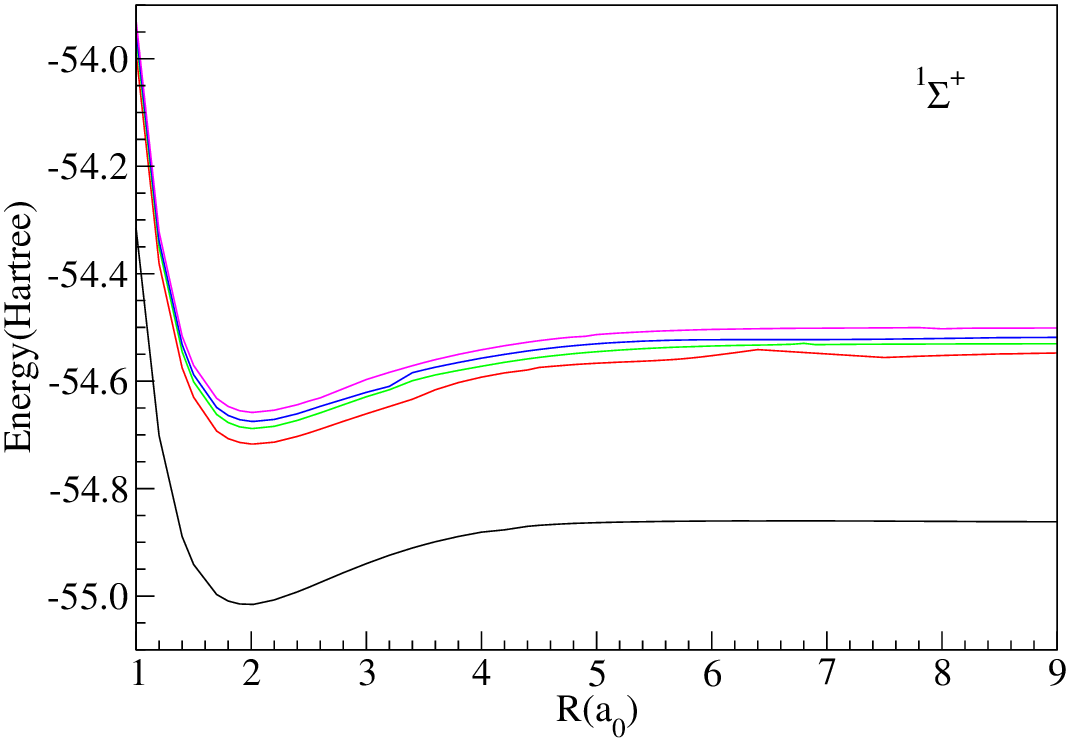}
\includegraphics[width=0.4\textwidth]{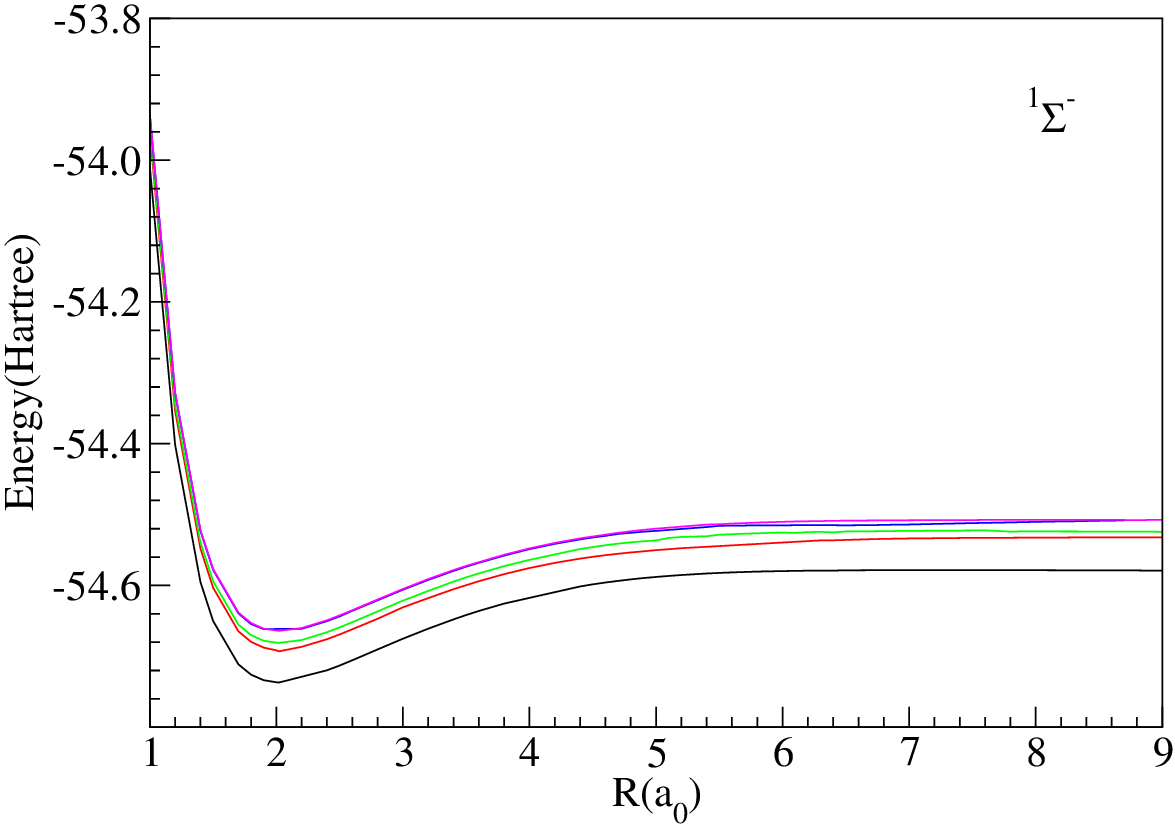}
\includegraphics[width=0.4\textwidth]{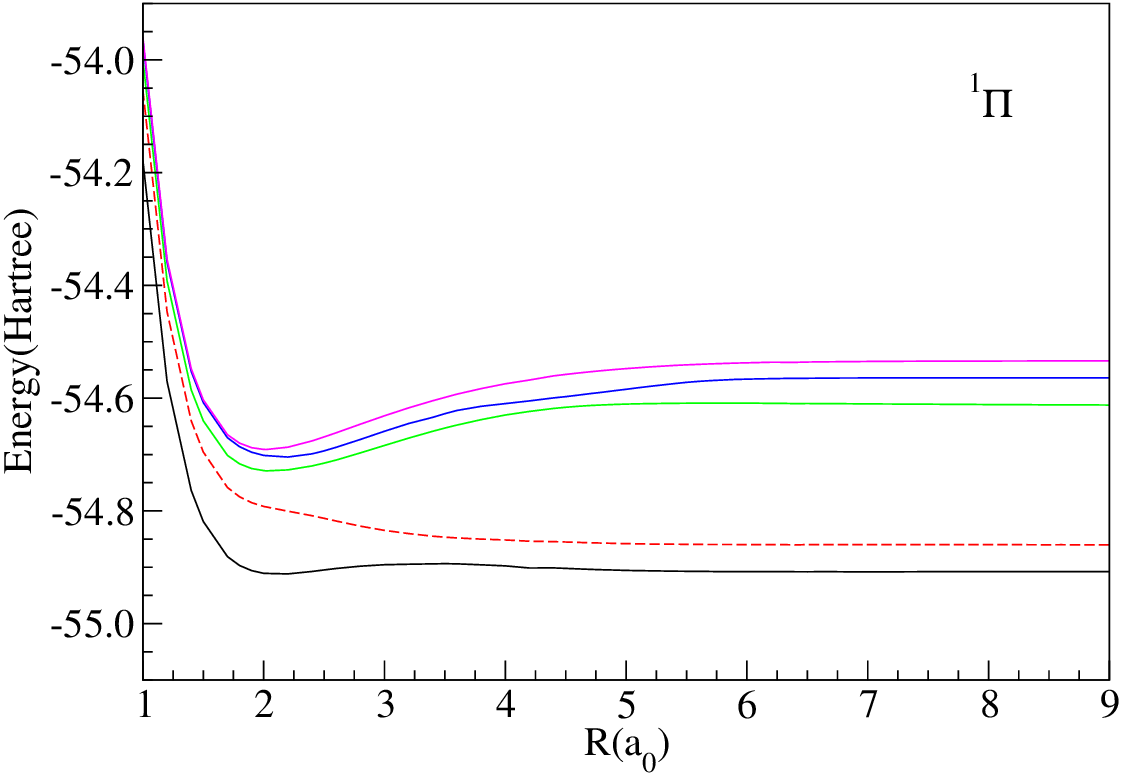}
\includegraphics[width=0.4\textwidth]{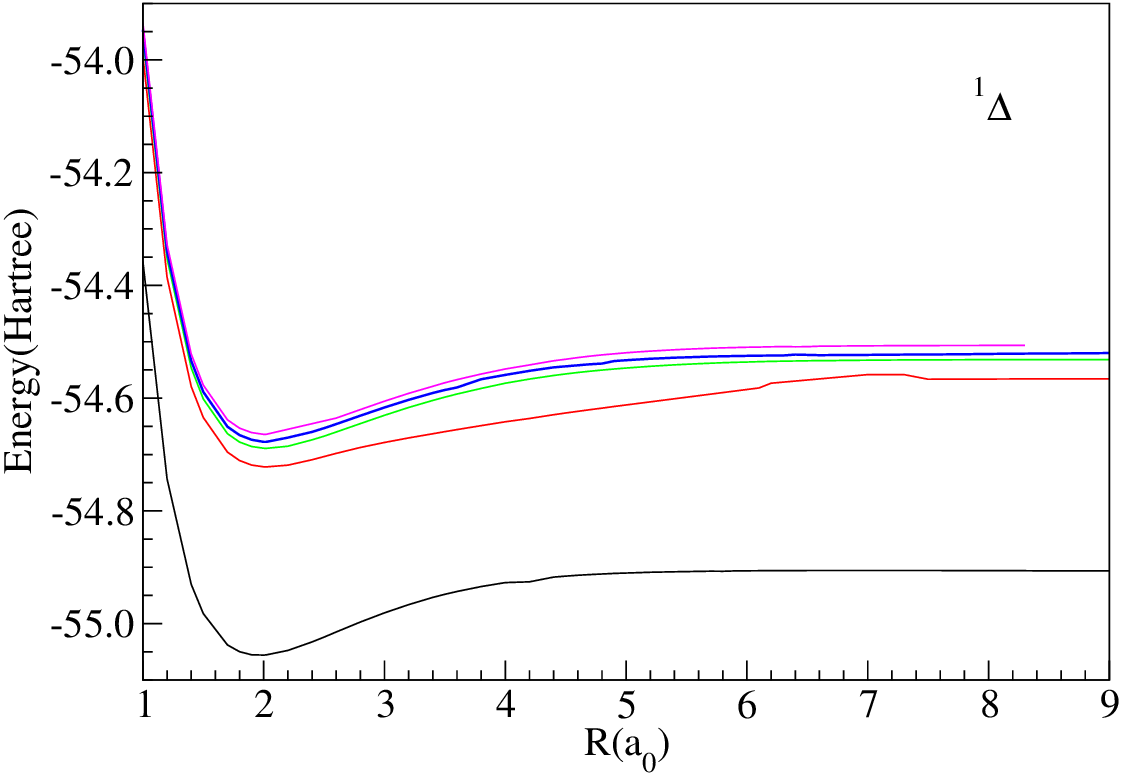}
\caption{Potential energy curves of the bound states of singlet symmetry of the NH molecule. The symmetry of the states are as indicated in each panel. The dashed curves are of valence character.\label{fig:bound1}}
\end{figure*}

\subsection{Identification of Rydberg states of NH at equilibrium}
In our previous work\cite{Ghosh22} we had calculated the VEEs from the X~$^3\Sigma^-$ ground state of the NH molecule to the first 29 excited states at its equilibrium geometry $R_e=2.0205$\bohr. The details of the PECs for many of these states were obtained by Owono \etal\cite{Owono07} where they focused mainly on the triplet and quintet states. The PECs for the singlet states were not presented by Owono \etal in as much detail as the triplets and quintets. Moreover, Owono \etal were also not able to clearly identify the Rydberg character of these states. There have been some earlier attempts to experimentally find new states and characterize their Rydberg orbitals, however, these again were also mainly limited to states with triplet symmetry\cite{Jhonson90,Clement92a,Clement92b}. 

We have tried to identify and list many of the Rydberg states of NH for singlet and triplet symmetries. In Table \ref{tab:QD} we have shown the Rydberg character of many of the states of NH at its equilibrium $R_e=2.0205$\bohr~characterized by their quantum defects. There is considerable disagreement regarding the nomenclature of the states between those of Owono \etal\cite{Owono07} and those of earlier works\cite{Jhonson90,Clement92a,Clement92b}. Therefore, to maintain consistency, for each symmetry we have numbered the states sequentially and have also kept the notations used in earlier works\cite{Owono07,Jhonson90,Clement92a,Clement92b} (in brackets) .

\subsubsection{Singlet states}
Subsequent to the work of Goldfield and Kirby\cite{Goldfield87} where they obtained PECs and dipole moments of the lowest $^3\Sigma^-$, $^3\Pi$ and $^5\Sigma^-$ symmetries, Johnson III and Hudgens\cite{Jhonson90} observed three new singlet Rydberg states, which they called the $f ^1\Pi(3p\sigma)$, $g ^1\Delta(3p\pi)$ and $h ^1\Sigma^+(3p\pi)$ states, by resonance enhanced multiphoton ionization (REMPI). These three states arise from the X~$^2\Pi$ ion core with a Rydberg electron in the states indicated in brackets. The quantum defects for these Rydberg orbitals can be calculated using the Rydberg formula 
\begin{equation}\label{eq:qd}
    T_0 = \mbox{IP} - Ry/(n-\mu)^2,
\end{equation}
where $T_0$ is the corresponding adiabatic excitation energy, IP the adiabatic ionization potential of NH, $Ry$ the Rydberg constant, $\mu$ the quantum defect and $n$ the principal quantum number of the Rydberg electron. 

\noindent Using the values of $T_0$ obtained by Johnson III and Hudgens\cite{Jhonson90} and the IP 13.49 eV quoted by them (taken from Lias et al.\cite{Lias88}), we obtained the quantum defects $\mu=0.788,\; 0.696,\; 0.614$ respectively for the Rydberg orbitals of the $f,\; g$ and $h$ states from Eq. (\ref{eq:qd}). From Table \ref{tab:QD}, our values of the quantum defect $\mu$ for the Rydberg orbitals (in the \Rmat formalism) corresponding to the $f,\; g$ and $h$ states are respectively $\mu = 0.718,\; 0.634,\; 0.566$ which are in good agreement with those of Johnson III and Hudgens. However, by analyzing the Rydberg series for $n > 3$ from the data available with us (in Table \ref{tab:QD}), we are inclined to label the $f$-state as $f^1\Pi(3s\sigma)$, rather than $f^1\Pi(3p\sigma)$ as proposed by Johnson III and Hudgens\cite{Jhonson90}. For the $b^1\Sigma^+(2p\pi)$ and $h^1\Sigma^+(3p\pi)$ states, the quantum defects of the Rydberg orbitals obtained from the parameters given by Johnson III and Hudgens as 0.881 and 0.614 respectively. These are in very good agreement with our calculated value 0.856 for the $2p\pi$ and 0.566 for the $3p\pi$ $^1\Sigma^+$ states. 

\begin{figure*}[t]
\centering
\includegraphics[width=0.4\textwidth]{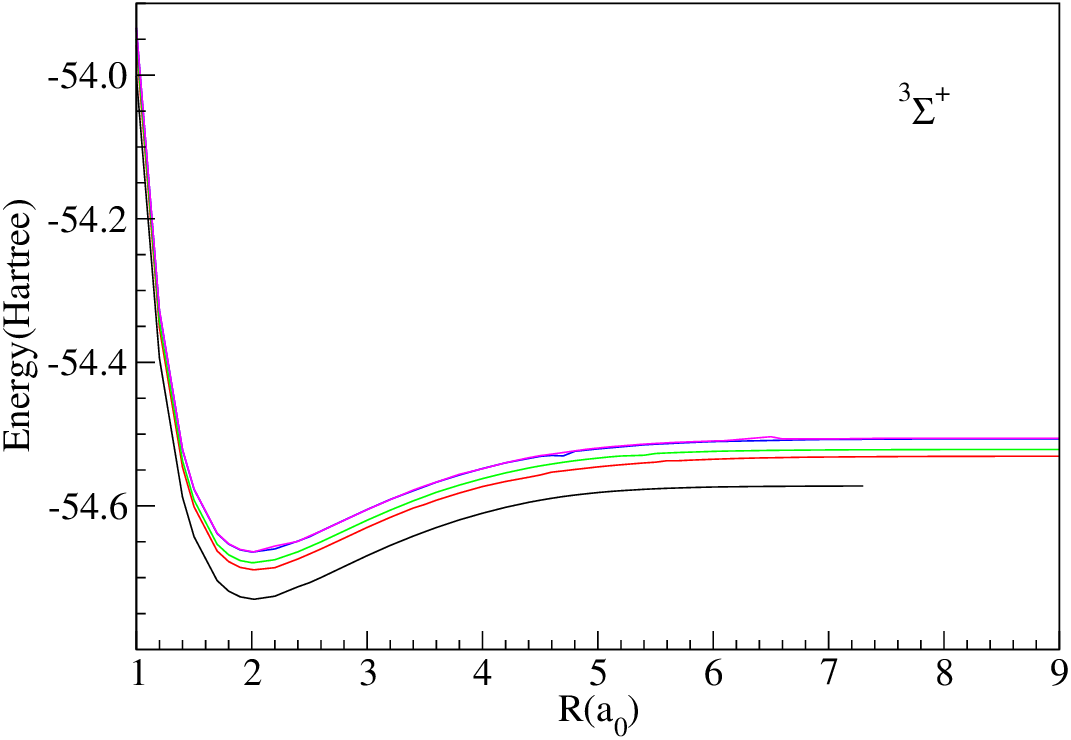}
\includegraphics[width=0.4\textwidth]{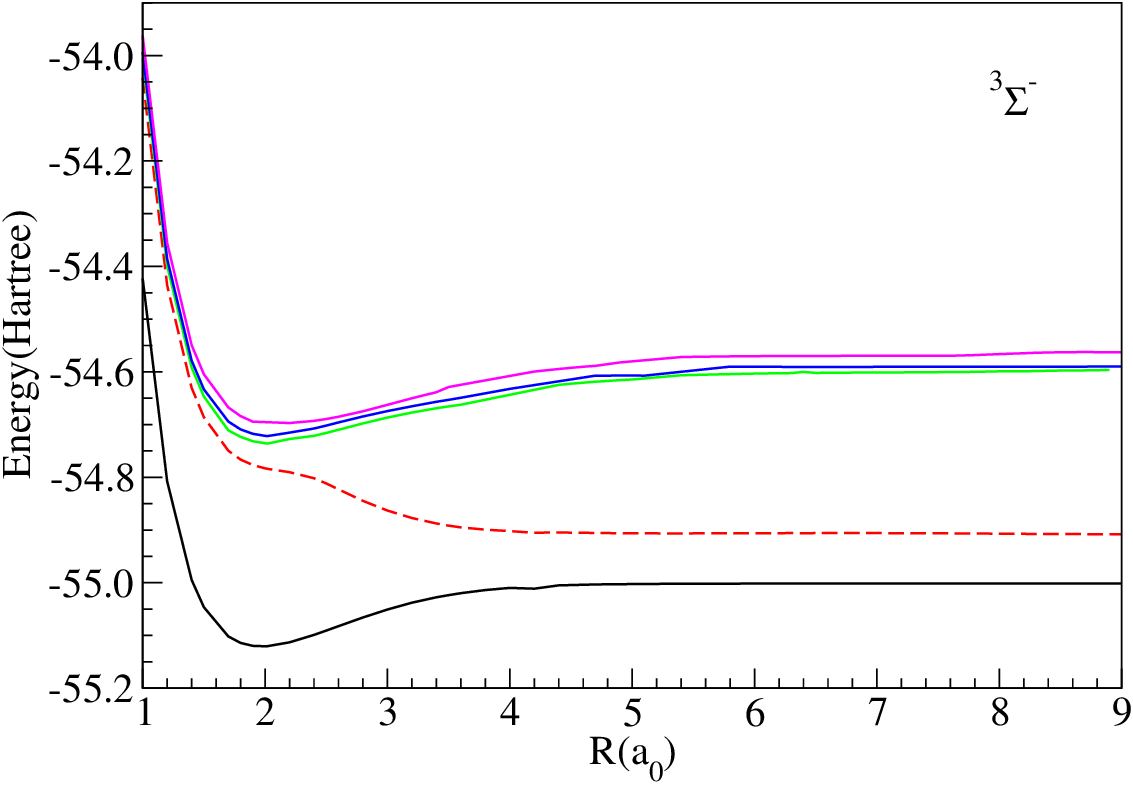}
\includegraphics[width=0.4\textwidth]{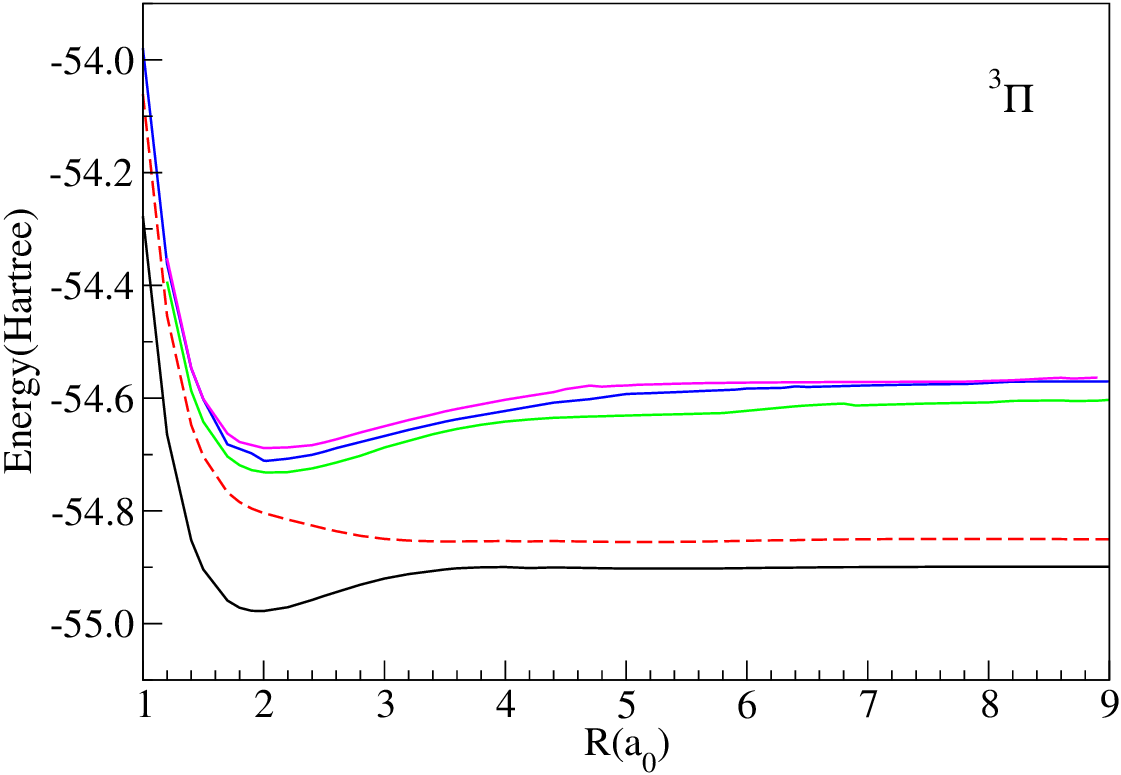}
\caption{Potential energy curves of the bound states of triplet symmetry of the NH molecule. The symmetry of the states are as indicated in each panel. The dashed curves are of valence character.\label{fig:bound3}}
\end{figure*}
\begin{figure*}[t]
\centering
\includegraphics[width=0.4\textwidth]{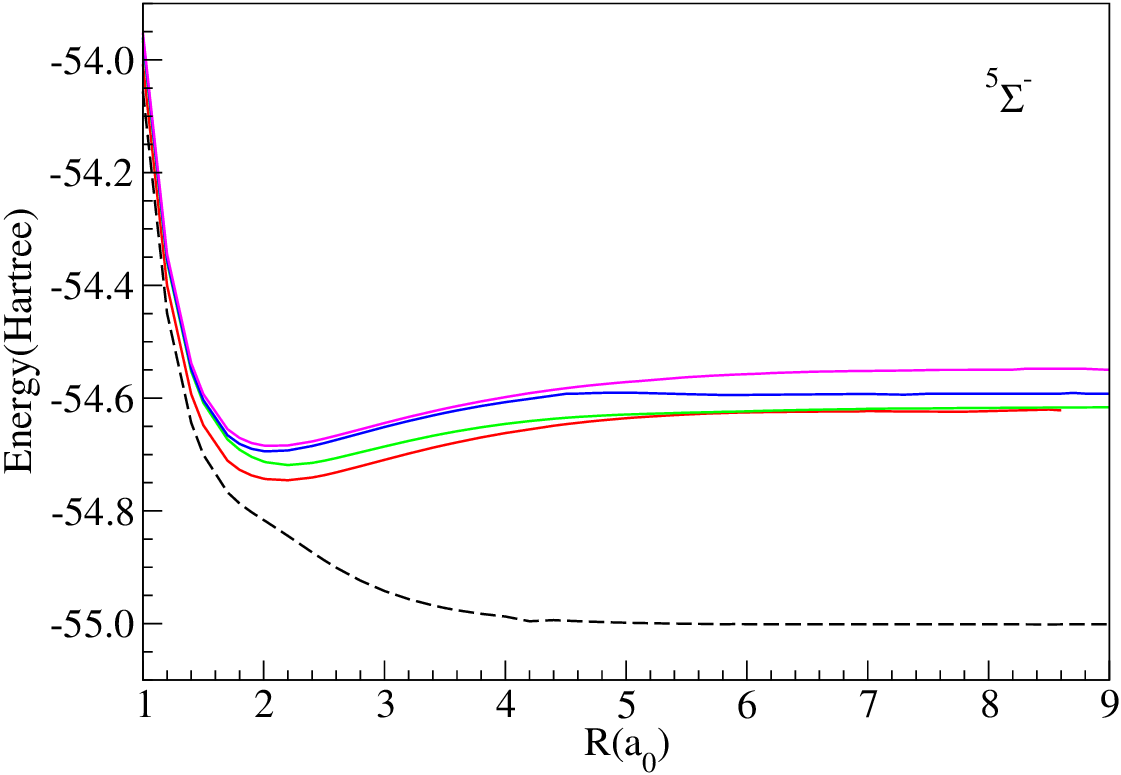}
\includegraphics[width=0.42\textwidth]{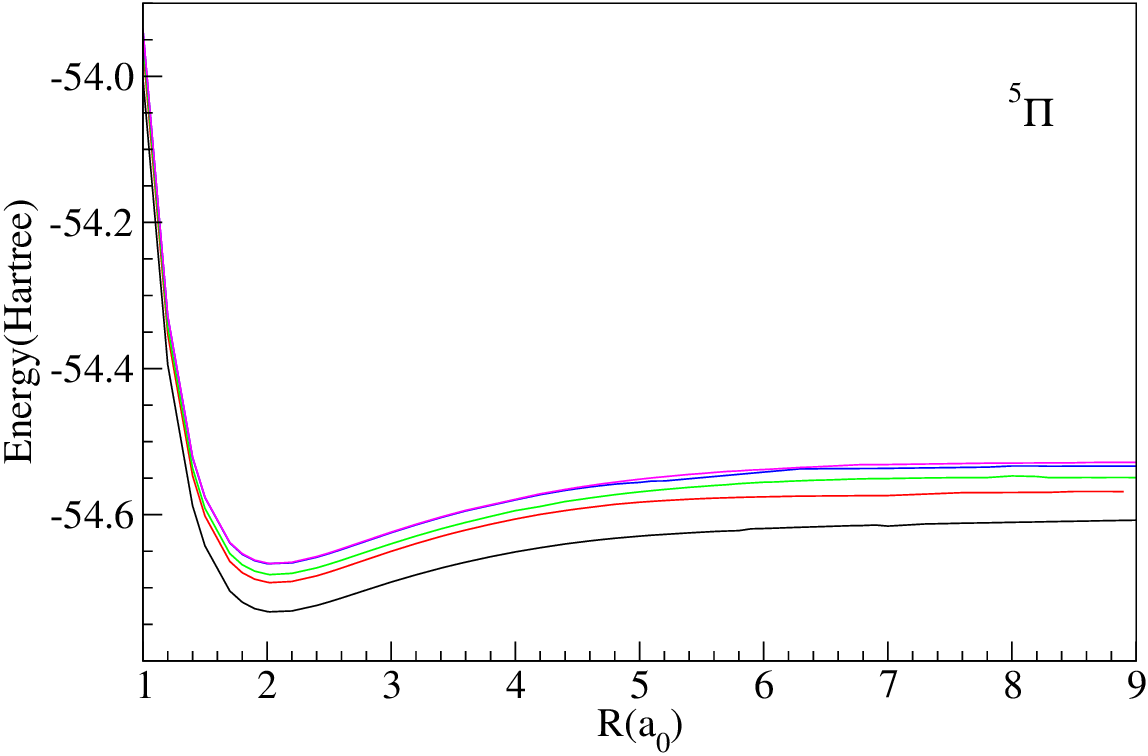}
\caption{Potential energy curves of the bound states of quintet symmetry of the NH molecule. The symmetry of the states are as indicated in each panel. The lowest $^5\Sigma^-$ (dashed) curve is of valence character.\label{fig:bound5}}
\end{figure*}

\subsubsection{Triplet states}
The $3~^3\Pi (3s\sigma)$ state is identified by  Clement \etal\cite{Clement92b} as the $B~^3\Pi$ state. From a calculated value of the quantum defect $\mu =0.820$ they showed that this state originates from the X~$^2\Pi$ ion core and a $3s\sigma$ Rydberg electron. The next $^3\Pi$ excited state, namely the $4~^3\Pi$, is identified by them as the $D~^3\Pi$ and is argued to originate from $a~^4\Sigma^-$ excited state, which lies very close to the ion ground state. From a calculated value of the quantum defect $\mu=0.640$, Clement \etal assigned to it a Rydberg orbital $3p\pi$. Our calculated values of the quantum defects for $3 (B)$ $^3\Pi$ and $4 (D)$ $^3\Pi$ states are respectively 0.727 and 0.440, which are slightly lower than the corresponding values quoted by Clement \etal\cite{Clement92b}. 

\noindent We attribute this to a different value of the ionization potentials used to calculate the quantum defects by Clement \etal The $5~^3\Pi$ state is not well documented and assuming $n=3$, we find a quantum defect $\mu = -0.059$. Since the immediately lower $B$ and $D$ states originate from the $a~^4\Sigma^-$ ion core, we believe this state also originates from the same ion core and hence we assign the Rydberg orbital $3d\sigma$ to it.

Overall, our obtained quantum defects are consistent and in good agreement with the experimentally determined values wherever available, and therefore the remaining ones are also likely to be reasonably accurate and consistent. We have also included the vertical excitation energies of most of the states in Table \ref{tab:QD} for reference and these are quoted from our previous work\cite{Ghosh22}.

\begin{figure*}[t]
\centering
\includegraphics[width=0.4\textwidth]{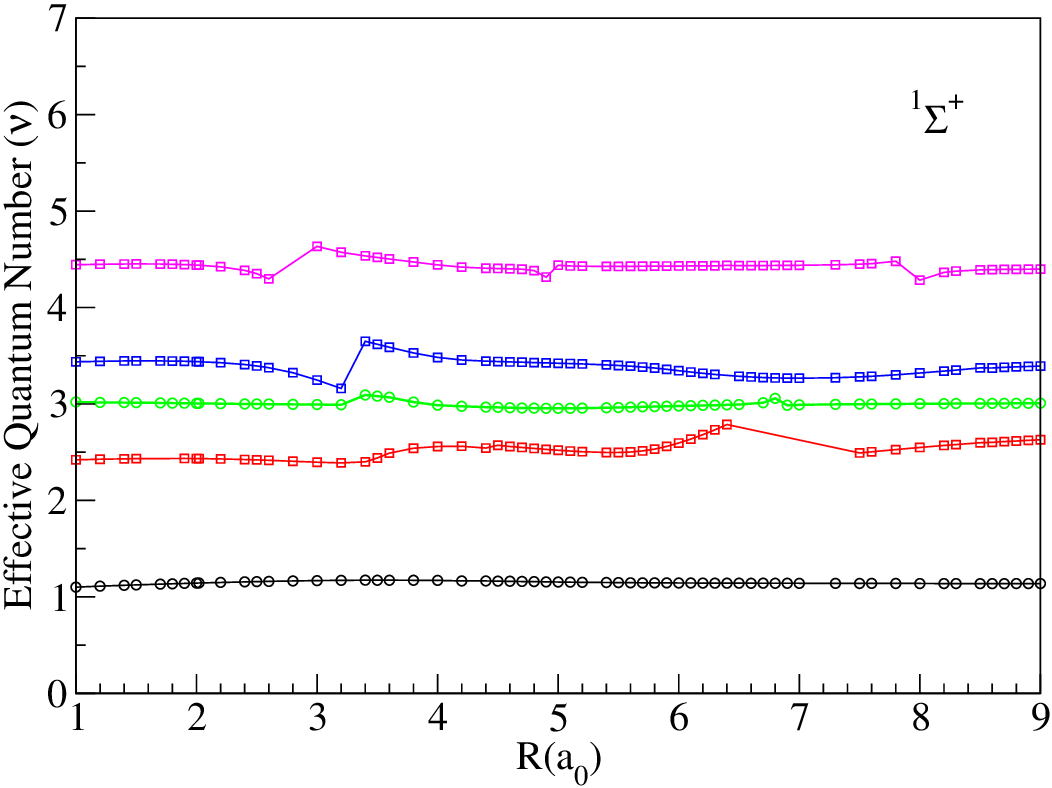}
\includegraphics[width=0.4\textwidth]{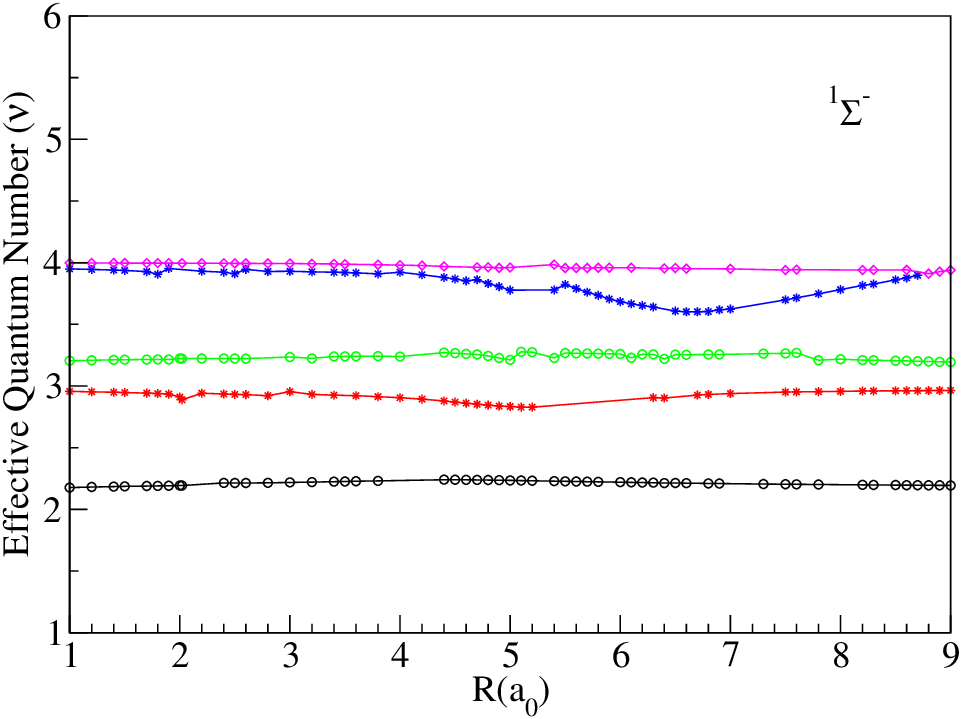}
\includegraphics[width=0.4\textwidth]{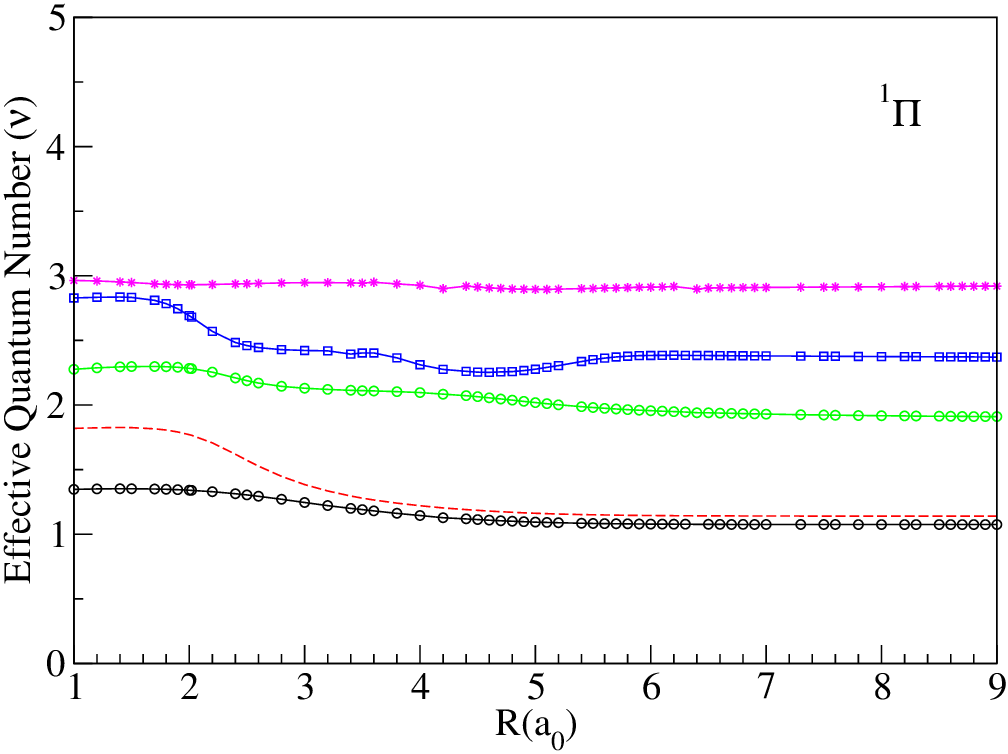}
\includegraphics[width=0.4\textwidth]{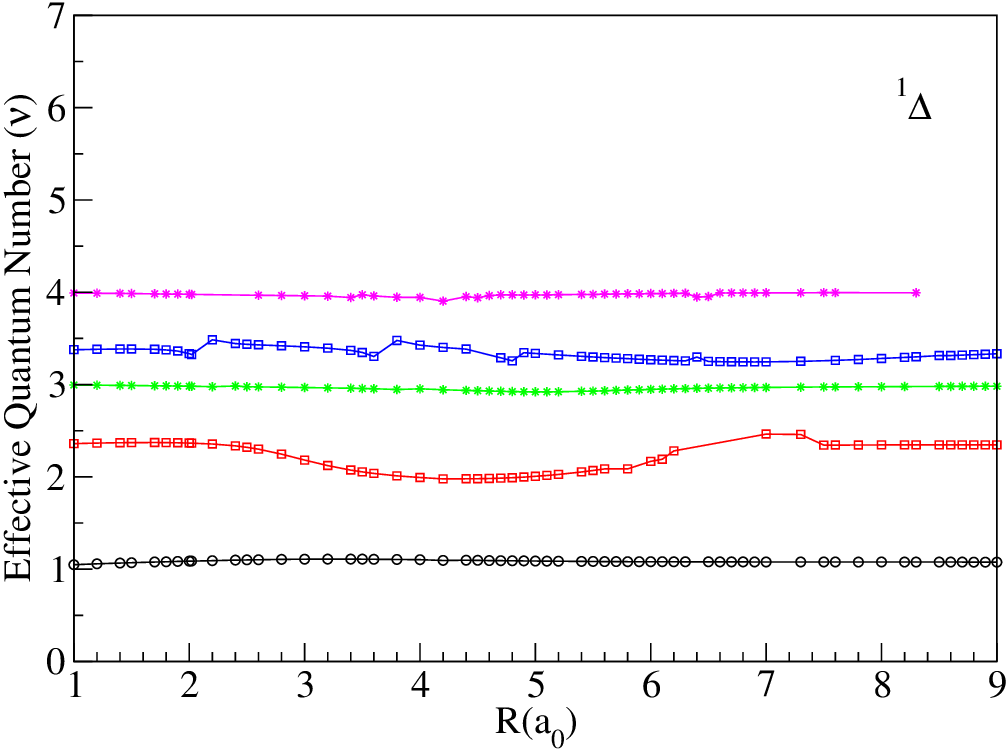}
\caption{Effective quantum numbers as a function of internuclear distance for some of the singlet bound states of the NH molecule. The symmetry of the states is indicated in each panel. Symbols: circle - s state, square - p state, star - d state, diamond - f state. Dashed curves correspond to valence states.}\label{fig:qd1}
\end{figure*}

\subsection{Potential energy curves for bound states of NH}

The relative positions of the first nine potential energy curves (PECs) are shown in Figure \ref{fig:bound}. The $^5\Sigma^-$ state is clearly of valence character, while the $^1\Sigma^-$ and the $^3\Sigma^+$ states appear to be valence for $R$ greater than 2.5\bohr. For a comparison, we have also shown the ground X~$^3\Sigma^-$ state computed by Owono \etal \cite{Owono07} (dashed curve in Figure \ref{fig:targ}) which agrees reasonably well with the X~$^3\Sigma^-$ state computed by us using the \Rmat formalism. 

Figures \ref{fig:bound1}, \ref{fig:bound3} and \ref{fig:bound5} show the potential energy curves (PECs) of some of the bound states of NH of singlet, triplet and quintet symmetries respectively. The PECs are adiabatic since our calculations are fixed nuclei calculations and hence, some of these, show signatures of avoided crossings.

Figures \ref{fig:bound1} shows the bound states of NH of singlet symmetries. The singlet symmetry means that all the states are coupled to the X~$^2\Pi$ ground state of the NH$^+$ ion. The 2~$^1\Pi$ (second $^1\Pi$ state from bottom shown as the dashed curve) appears to be valence for $R > 2.0$\bohr. Some of the curves display undulations due to avoided crossings.

Figures \ref{fig:bound3} and \ref{fig:bound5} show the PECs of some of the bound states of NH of triplet and quintet symmetries. These either arise from the X~$^2\Pi$ ground state of NH$^+$, or its $a~^4\Sigma^-$ excited state which lies very close to the ground state. The second $^3\Sigma^-$ and $^3\Pi$ states are likely to be valence for $R > 2.0$\bohr, as is the lowest $^5\Sigma^-$ state.
The effective quantum numbers $\nu$, defined as $\nu = n - \mu$, $\mu$ being the quantum defect, when plotted as a function of internuclear distance $R$ gives a more complete assessment of the number and character of states of each symmetry, than by simply looking at their PECs. The effective quantum numbers are also more sensitive to change of the internuclear distance $R$ than the PECs and hence, are better candidates for identifying adiabatic effects such as avoided crossings. Additionally, such plots also help in identifying, what are called the intruder states, ie. bound states which do not belong to any Rydberg series.

\begin{figure*}[t]
\centering
\includegraphics[width=0.4\textwidth]{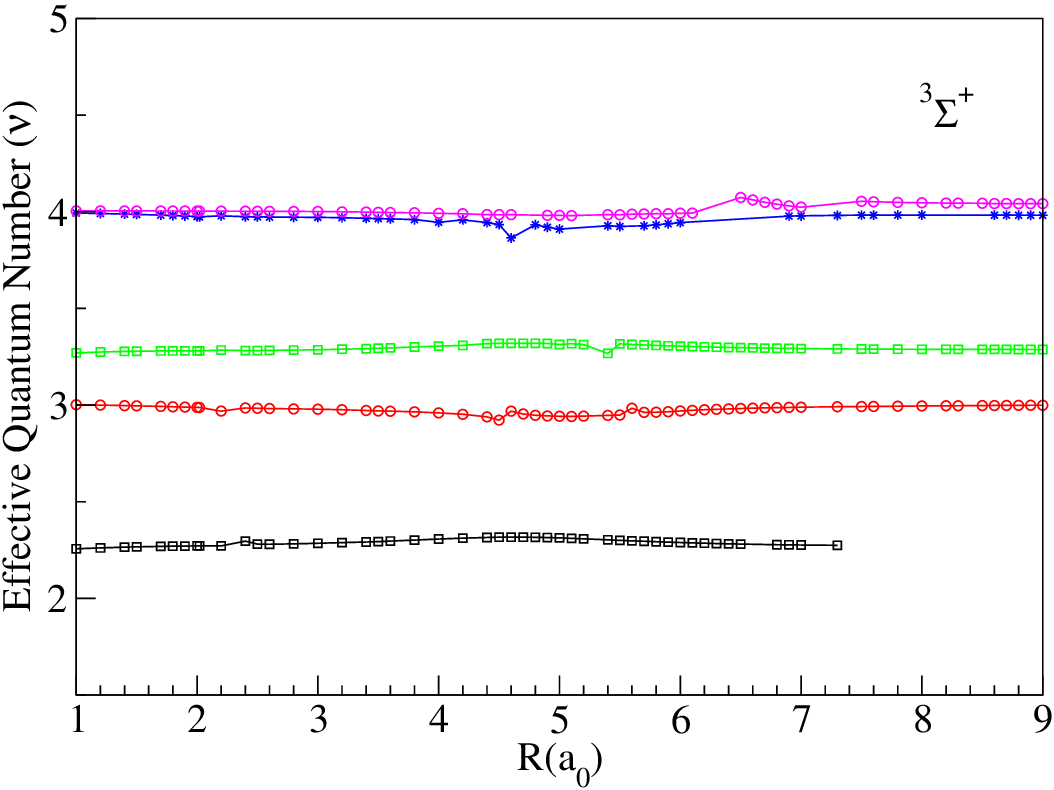}
\includegraphics[width=0.4\textwidth]{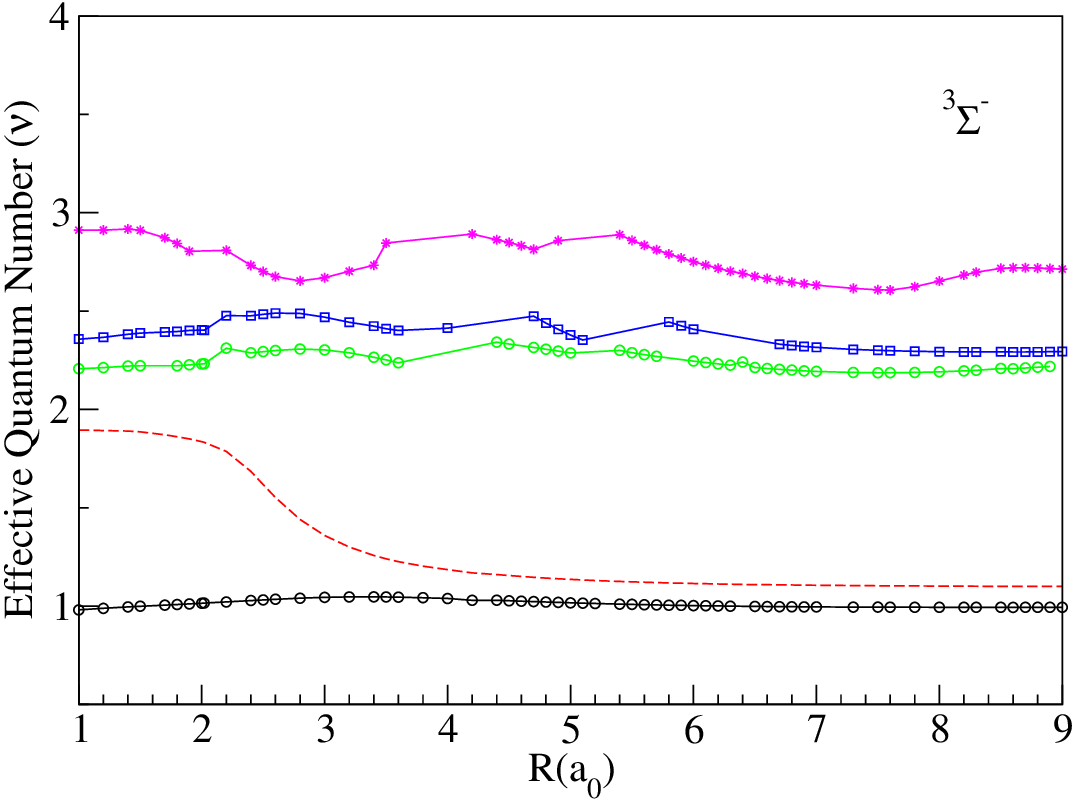}
\includegraphics[width=0.4\textwidth]{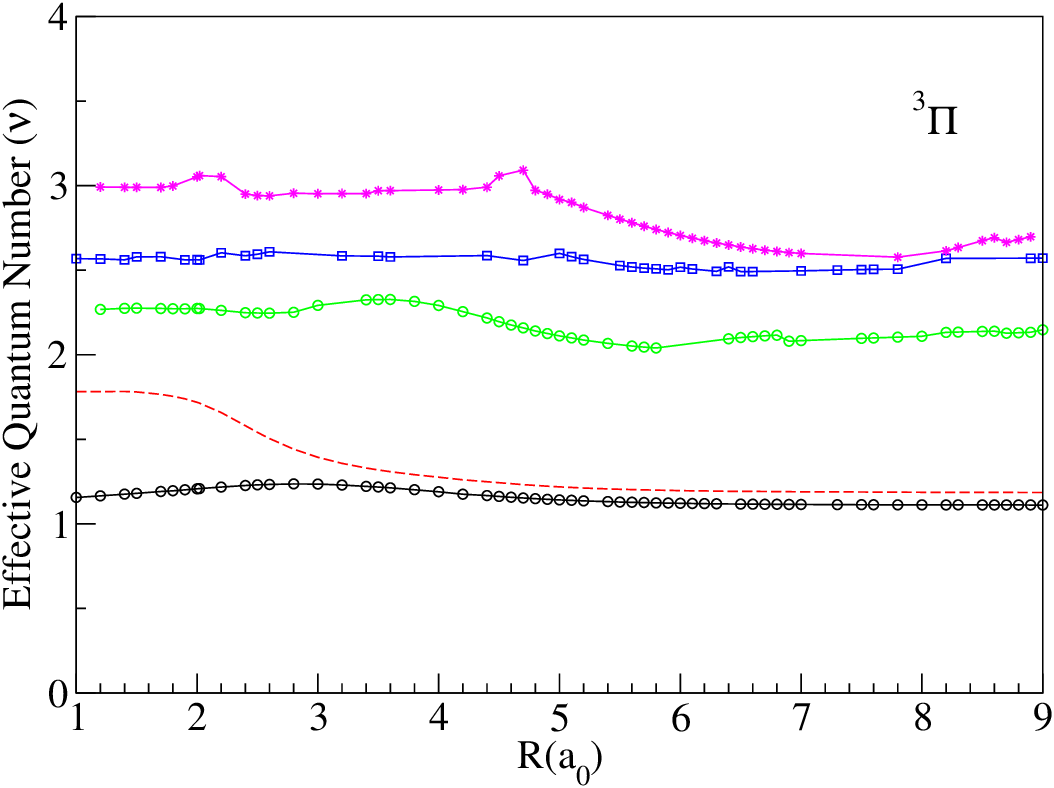}
\caption{Effective quantum numbers as a function of internuclear distance for some of the triplet bound states of the NH molecule. The symmetry of the states is indicated in each panel. Symbols: circle - s state, square - p state, star - d state, diamond - f state. Dashed curves correspond to valence states.}\label{fig:qd3}
\end{figure*}
\begin{figure*}[t]
\centering
\includegraphics[width=0.4\textwidth]{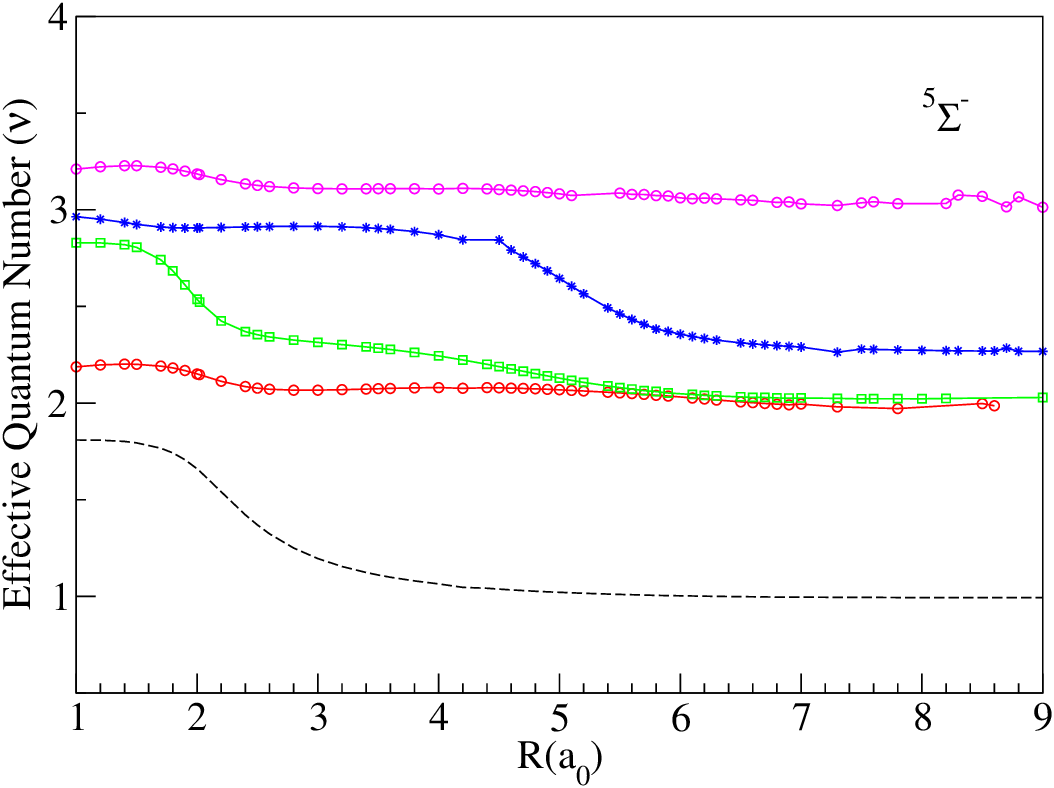}
\includegraphics[width=0.4\textwidth]{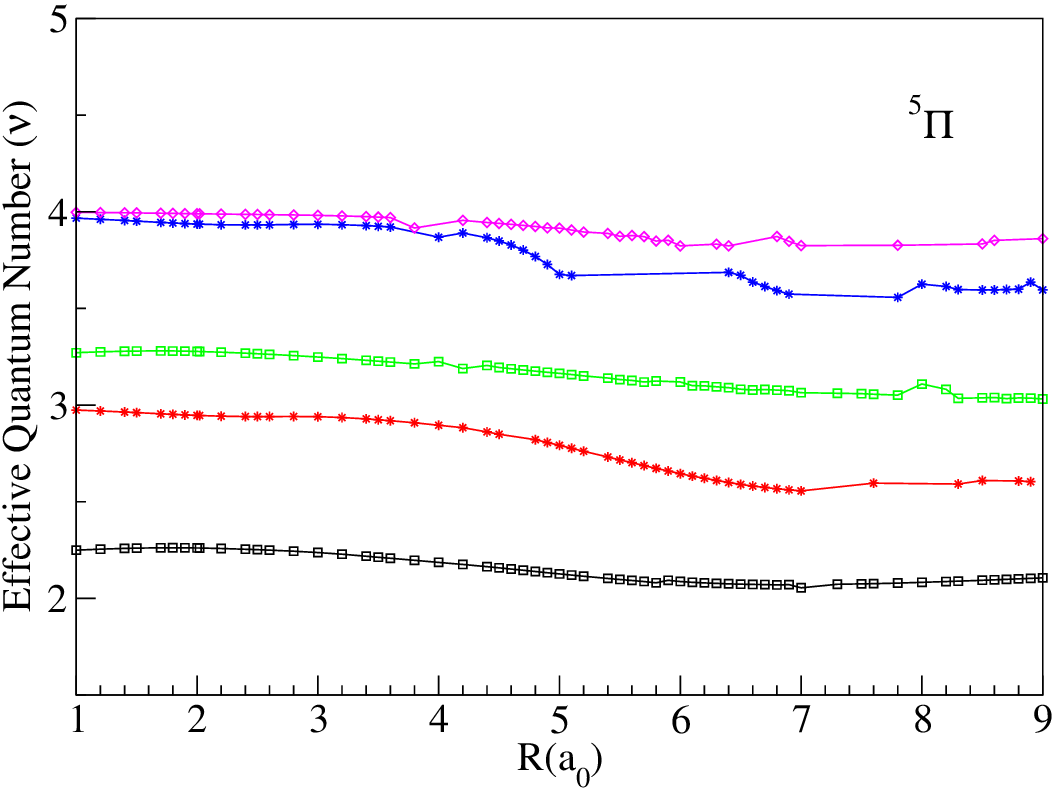}
\caption{Effective quantum numbers as a function of internuclear distance for some of the quintet bound states of the NH molecule. The symmetry of the states is indicated in each panel. Symbols: circle - s state, square - p state, star - d state, diamond - f state. Dashed curves correspond to valence states.}\label{fig:qd5}
\end{figure*}

The intruders often are the continuation of resonant states situated above the ion as bound states lying below the ion,  and are therefore parts of full diabatic states. As is well known, such diabatic states play a very important role in many electron driven dissociating processes, for example, dissociative recombination (DR) and resonant dissociative excitation (RDE). One of the goals of this work is to identify such diabatic states and provide corresponding Rydberg-valence couplings, or equivalently, resonance widths, that would be useful for further DR studies on the NH$^+$ ion.

Figures \ref{fig:qd1}, \ref{fig:qd3} and \ref{fig:qd5} show the effective quantum numbers as a function of internuclear distance $R$ for the Rydberg states of singlet, triplet and quintet character respectively. Many of the curves display sharp kinks, particularly for high $\nu$, due to avoided crossings. Further, the second (from bottom) $^1\Pi$, $^3\Sigma^-$ and $^3\Pi$ states and the lowest $^5\Sigma^-$ state, shown as dashed curves, appear to be predominantly of valence character. 

\begin{figure*}[t]
\centering
\includegraphics[width=0.44\textwidth]{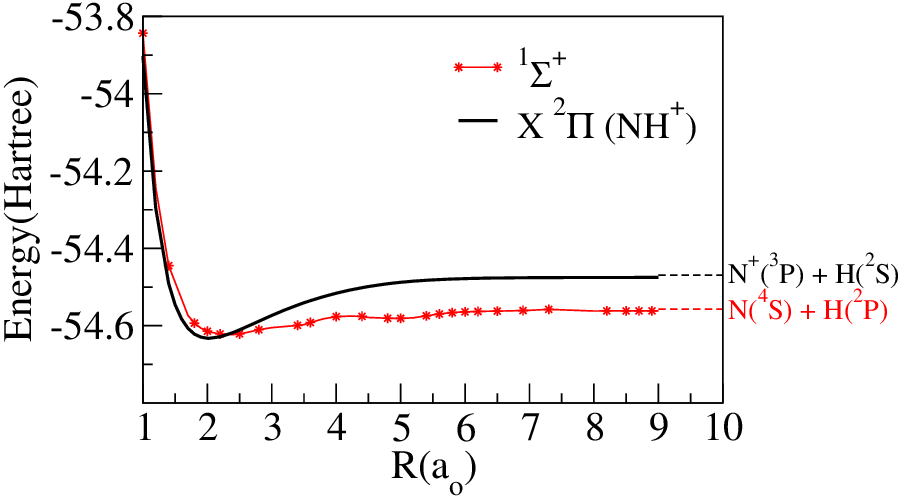}
\includegraphics[width=0.44\textwidth]{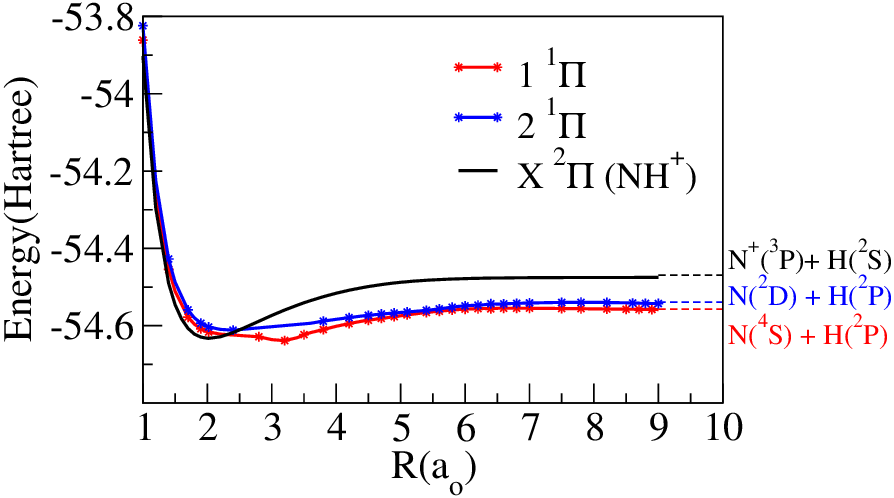}
\includegraphics[width=0.44\textwidth]{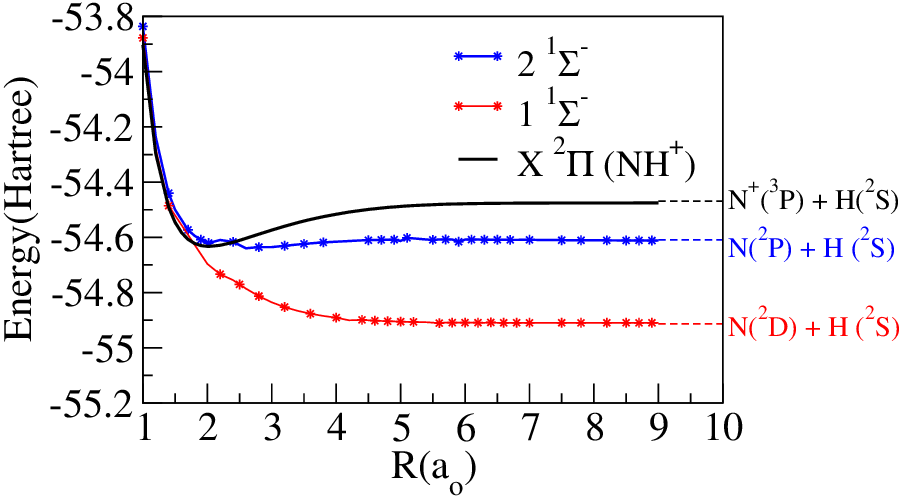}
\includegraphics[width=0.44\textwidth]{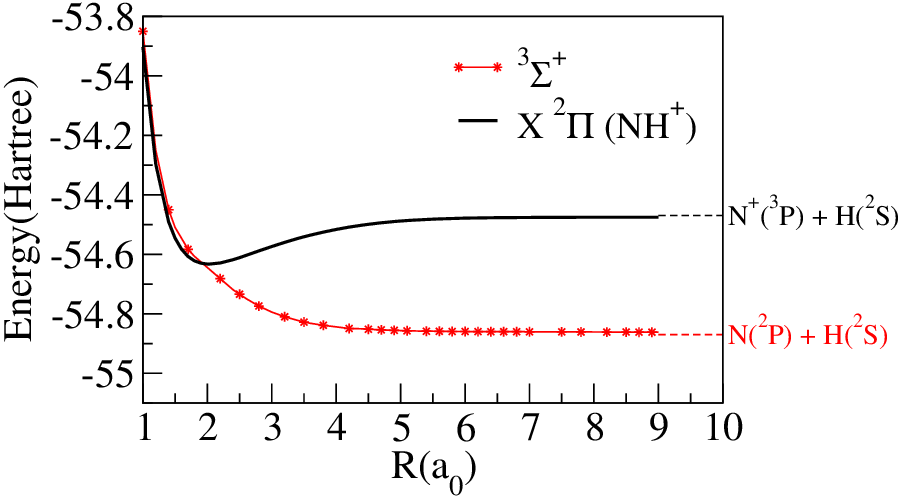}
\caption{Lines with symbols: Resonance curves of symmetries indicated in the legend of each panel. Continuous curve: NH$^+$ ion X~$^2\Pi$ ground state. Resonances become bound after crossing the ion from above.}\label{fig:res}
\end{figure*}
\begin{figure*}[t]
\centering
\includegraphics[width=0.44\textwidth]{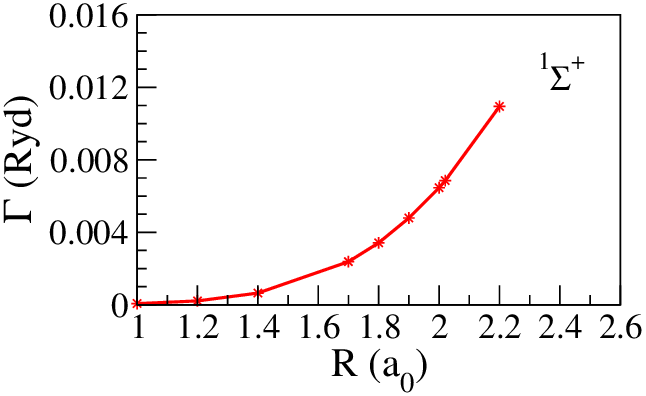}
\includegraphics[width=0.44\textwidth]{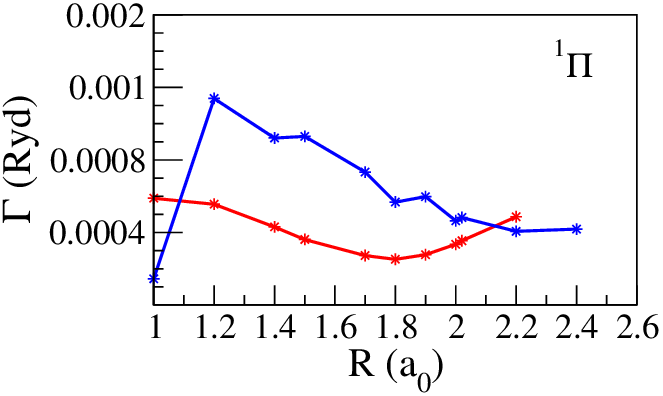}
\includegraphics[width=0.44\textwidth]{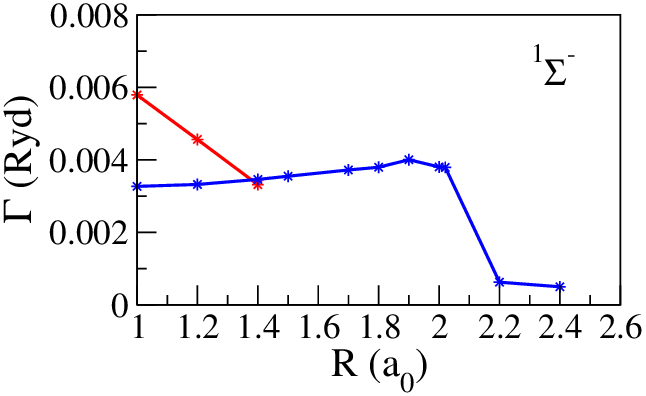}
\includegraphics[width=0.44\textwidth]{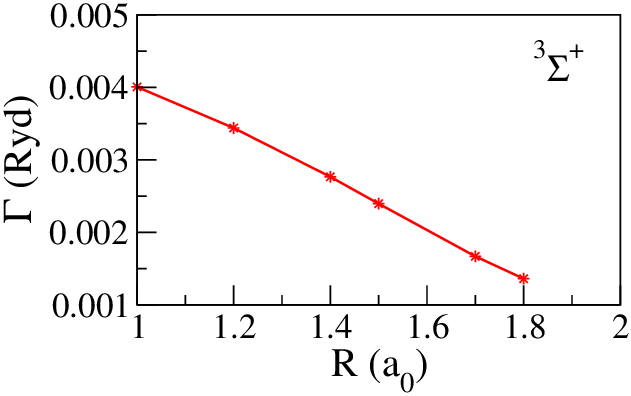}
\caption{Lines with symbols: Resonance widths corresponding to the resonance curves and with the same colour coding shown in Figure \ref{fig:res}. Symmetry of the states is indicated in each panel.}\label{fig:reswidth}
\end{figure*}

\subsection{Resonance curves and widths}
Resonances play a very important role in many electron collision processes in molecular ions such as dissociative recombination which undergo via neutral resonant states and lead to release of neutral fragments. In the adiabatic, fixed nuclei calculation which we adopt here, the full resonance curves correspond to neutral dissociative states in a diabatic picture. The resonant states have different behaviour above and below the ion (here NH$^+$). Above the ion, the neutral resonant states are coupled to the ionisation continua and hence can quickly disintegrate by auto ionization into an ion and an electron. The resonant states therefore have finite lifetimes $\tau$ that vary inversely as the resonance width $\Gamma$. Below the ion the resonances become bound and have widths $\Gamma=0$, and therefore have infinite lifetimes and are thus electronically stable.

In our previous work \cite{Ghosh22} we had obtained the resonant states of the e+NH$^+$ system at a single geometry, namely the NH$^+$ equilibrium $R_e=2.0205$\bohr~categorized by their parent state. In this work, we present a comprehensive set of resonance curves and their widths as a function of inter nuclear distance, none of which were reported before.  For the case of NH$^+$, the neutral resonant states are pathways for collision processes like DR that release atomic nitrogen. Such collision processes are therefore of great significance in fusion tokamaks where nitrogen seeding is used to mitigate heat load on the plasma facing components. The present molecular data is therefore expected to be of considerable importance as primary inputs relevant theoretical calculations.

Figure \ref{fig:res} shows the resonance potential energy curves of $^1\Sigma^+$, $^1\Pi$ symmetries (top panels) and $^1\Sigma^-$, $^3\Sigma^+$ symmetries (bottom panels). The dissociation limits and dissociation energies of these states are summarized in Table \ref{tab:disslim}. Most of the PECs of these states pass very close to the Franck-Condon region near the NH$^+$ equilibrium. It is known that the direct DR cross sections are proportional to the square of the overlap between the wave functions of the vibrational state of the ion \cite{Djuissi24} and the resonant dissociative state, and such overlaps are strongest for the ground vibrational state $\xi_0$ of the ion and a resonant state, if the resonant state has a good Franck-Condon overlap with the ground state of the target. 
In this respect, the resonant states we have obtained here are likely to be strong candidates for DR pathways at low collision energies. In Figure \ref{fig:reswidth} we have also shown the global resonance widths corresponding to the resonant states shown in Figure \ref{fig:res}. The widths $\Gamma^r(R)$ corresponding to the resonances are small and are related to the Rydberg-valence couplings $V_{el}^r(R)$, which represent the couplings of these states to the ionization continuua, through the relation $V_{el}^r(R)=\sqrt{\Gamma^r(R)/2\pi}$ and it is the driving force of DR. A proper validation of our molecular data is not possible in the absence of other similar results. However, the quality of these can be assessed once DR cross sections calculations are performed with these data and compared with experimental DR cross sections which are already available\cite{Novotny2014,Yang2014,Lecointre10}.

\begin{table}[t]
\caption{Dissociation limits and dissociation energies (in Hartree) for the resonant states shown in Figure \ref{fig:res}.}

\label{tab:disslim}
\begin{tabular}{l l l} 
 \hline
States  & Dissociation & Dissociation\\
        & limits       & energies$^*$\\
 \hline
 2~$^1\Pi$ & N($^2$D)+H($^2$P) & -54.539\\
 $^1\Sigma^+$, 1~$^1\Pi$ & N($^4$S)+H($^2$P) & -54.557\\
 2~$^1\Sigma^-$, $^3\Sigma^+$ &N($^2$P)+H($^2$S) & -54.609 \\
 $^3\Sigma^+$ &N($^2$P)+H($^2$S) & -54.870 \\
 1~$^1\Sigma^-$ &  N($^2$D)+H($^2$S) & -54.914\\
\hline
\end{tabular}\\
$^*$Dissociation energies have been calculated using data from NIST. https://www.nist.gov/pml/atomic-spectra-database.
\end{table}

\section{Conclusions}
With the aim of identifying Rydberg states and resonant states of NH, in the present work we have undertaken detailed \Rmat calculations. Although many of the potential energy curves for the bound NH states have been obtained earlier by Owono \etal \cite{Owono07}, we have clearly identified many of these as Rydbergs by analyzing the quantum defects of the corresponding excited Rydberg electron, and have categorized them into Rydberg series according to their orbital quantum numbers. Finer details on the variation of the effective quantum numbers of the Rydberg states with the internuclear distance $R$, which provides deeper insights into their nature than the PECs are also provided and some of the valence states have been identified. We have also made detailed calculations on the resonant states and the resonance widths, and their continuation as bound states below the target NH$^+$ ion. These have been used to derive complete diabatic states of $^1\Sigma^+, ^1\Pi, ^1\Sigma^-$ and $^3\Sigma^+$ symmetries. The details of the Rydberg states, the resonant states, their widths and the full diabatized curves which serve as primary data for processes like dissociative recombination, dissociative excitation, resonant vibrational excitation and de-excitation have been obtained here for the first time. The absence of molecular data on the resonant states and their widths is principally the reason for lack of progress on theoretical DR studies on NH$^+$, despite some experimental work in this direction\cite{Novotny2014,Yang2014,Lecointre10}, and therefore, we believe the molecular data from the present work would lead to significant advancements in the electron driven reactive processes 
that undergo via these resonant states.

\section*{Author Contributions}
All authors contributed equally to this work.

\section*{Conflicts of interest}
There are no conflicts to declare.

\section*{Acknowledgments}
This work was supported by the SERB, New Delhi, under the Core Research Grant CRG/2021/000357 provided to KC. 

JZsM is grateful for financial support from the National Research, Development, and Innovation Fund of Hungary, under Project Number FK 132989. 

IFS acknowledges 
support from F\'ed\'eration de
Recherche par Confinement Magn\'etique (CNRS and
CEA), La R\'egion Normandie, LabEx EMC3 (project PTOLEMEE), COMUE Normandie 
Universit\'e,  
Institute for Energy, Propulsion and Environment, International Atomic Energy 
Agency via the project "The Formation and Properties of Molecules in Edge Plasmas",  
Agence Nationale de la Recherche via the project MONA, and
Programme National “Physique et Chimie du Milieu Interstellaire” of CNRS/INSU with 
INC/INP co-funded by CEA and CNES. 

NP thanks the International Atomic Energy Agency (IAEA) via the project CRP: ‘‘The Formation and Properties of Molecules in Edge Plasmas’’ and the European Union via COST (European Cooperation in Science and Technology) actions: PLANETS (CA22133), PROBONO (CA21128), PhoBioS (CA21159), COSY (CA21101), AttoChem (CA18222) and DYNALIFE (CA21169).
\section*{Data availability}
Upon a reasonable request, the data supporting this article will be provided by the corresponding author.

\bibliographystyle{unsrt}
\bibliography{NH,NHp}

\end{document}